\documentclass[12pt, a4paper]{article}
\usepackage[font=footnotesize]{caption}
\usepackage[utf8]{inputenc}
\usepackage{graphicx}
\usepackage{authblk}
\usepackage{booktabs}
\usepackage{verbatim}
\usepackage{url}
\usepackage{caption}
\usepackage{subcaption}
\usepackage{indentfirst}
\usepackage{url}
\usepackage{amssymb}
\usepackage{amsmath}
\usepackage{tcolorbox}
\title{Subsuming Complex Networks \\ by Node Walks}
\author{Alexandre Benatti$^1$ \\ Luciano da F. Costa$^2$}

\affil{
$^1$Institute of Mathematics and Statistics - DCC \\
University of S\~ao Paulo \\
Rua do Mat\~ao, 1010, \\ S\~ao Paulo, SP 05508-090 Brazil 
\\ \vspace{0.5cm}
$^2$S\~ao Carlos Institute of Physics - DFCM \\
University of S\~ao Paulo \\
Av. Trabalhador S\~ao-carlense, 400, \\ S\~ao Carlos, SP 13566-590 Brazil
}

\date{\emph{22th March, 2024}}

\begin{document}

\maketitle

\begin{abstract}
The concept of node walk in graphs and complex networks has been addressed, consisting of one or more nodes that move into adjacent nodes, henceforth incorporating the respective connections. This type of dynamics is then applied to subsume complex networks. Three types of networks (Erd\H{o}s- R\'eny, Barab\'asi-Albert, as well as a geometric model) are considered, while three node walks heuristics (uniformly random, largest degree, and smallest degree) are taken into account. Several interesting results are obtained and described, including the identification that the subsuming dynamics depend strongly on both the specific topology of the networks as well as the criteria controlling the node walks. The use of node walks as a model for studying the relationship between network topology and dynamics is motivated by this result. In addition, relatively high correlations between the initial node degree and the accumulated strength of the walking node were observed for some combinations of network types and dynamic rules, allowing some of the properties of the subsumption to be roughly predicted from the initial topology around the waking node which has been found, however, not to be enough for full determination of the subsumption dynamics. Another interesting result regards the quite distinct signatures (along the iterations) of walking node strengths obtained for the several considered combinations of network type and subsumption rules.
\end{abstract}

\section{Introduction}\label{sec:introduction}

Graphs and complex networks have undergone impressive scientific-tech\-nolo\-gi\-cal development along the last decades mainly as a consequence of their ability to represent virtually every discrete system (e.g.~\cite{da2018complex}). Basically, complex networks can be informally defined as graphs with a more elaborate topology than uniform graphs, such as regular graphs (where all nodes have the same degree) and stochastically regular networks (where most nodes have similar degree values).

In addition to their intrinsic ability to represent discrete systems, graphs, and complex networks can also be effectively used to model complex dynamical systems, including those in which the topology of the network changes along time (e.g.~\cite{tranos2015international,hofmann2018complex,silva2022self}) while being potentially influenced by the respective dynamics performed at the level of dynamical states associated to the nodes. These combined possibilities make complex networks a primary resource to be considered for scientific modeling (e.g.~\cite{da2019modeling}) not only of a wide range of real-world systems but also a virtually unlimited number of possible complex systems involving intricate topology and dynamics.

\begin{figure}
  \centering
     \includegraphics[width=.95 \textwidth]{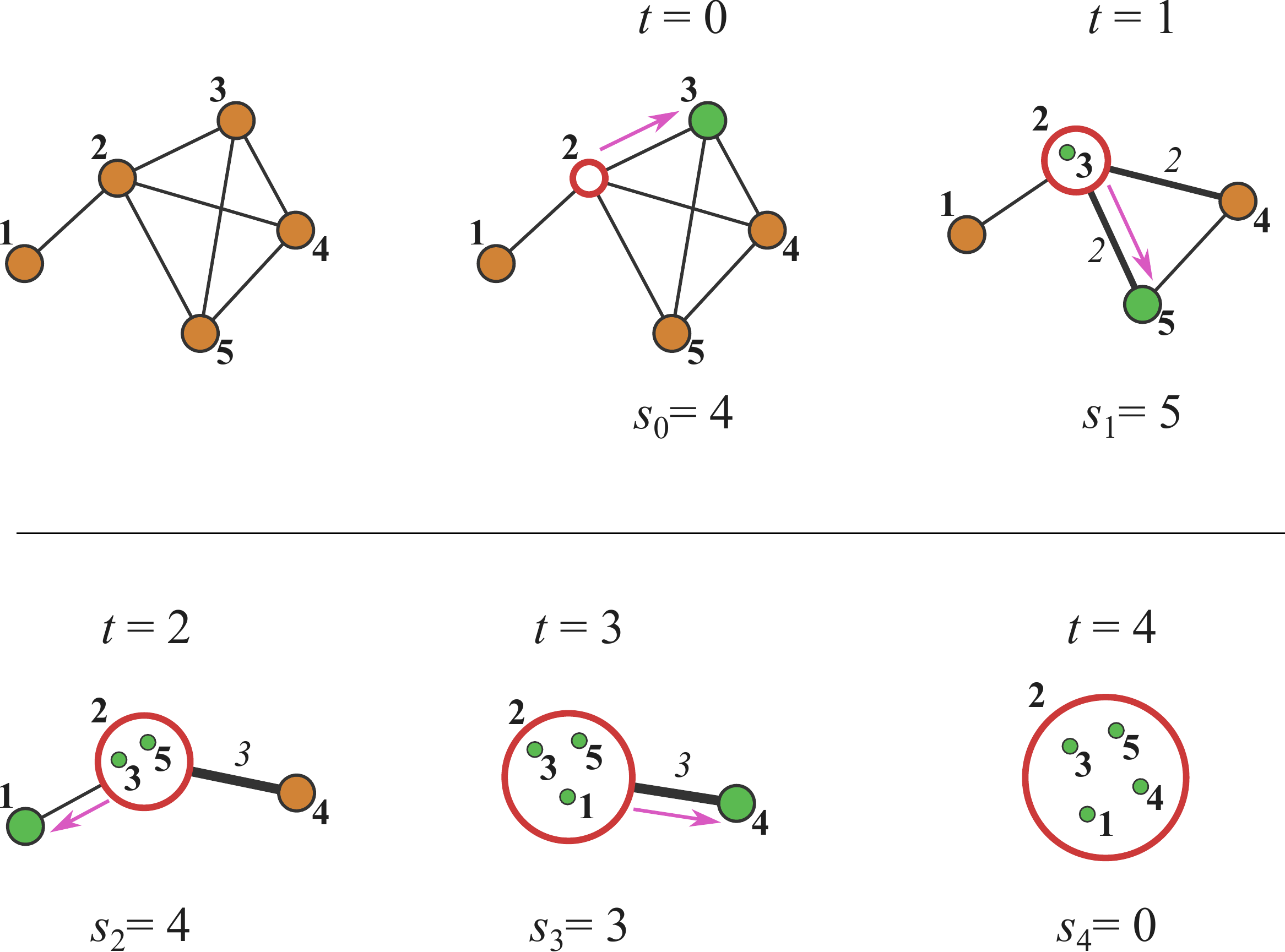}
   \caption{The subsumption of a graph performed along iterations $t=0, 1, 2, 3, 4$ respectively to walking node 2 (shown with red border) and choosing the next node (in green) in a uniformly random manner. At each iteration, one of the neighbors of the walking node is chosen with uniform probability and incorporated into the walking node, which inherits the respective links. Subsumed links have their links superimposed. For instance, the incorporation of node 3 into the walking node 2 leads to the links to nodes 4 and 5 to have increased from 1 to 2. The strength $s$ of the walking node, corresponding to the sum of the weights of the respective links, is also shown for each iteration.}\label{fig:walk}
\end{figure}

The present work aims at addressing an intricate situation, involving networks that undergo successive topological changes --- henceforth called \emph{subsumptions} --- as a consequence of random walks being performed by one or more of its nodes, henceforth referred to as \emph{walking nodes} (please refer to Fig.~\ref{fig:walk}). The obtained walks are henceforth understood as \emph{node walks}, in order to reflect the fact that the walking node successively displaces itself onto the selected nodes to be subsumed.

More specifically, given a network and one of its nodes as a walking node, one of its neighbors is chosen according to an underlying rule, with these two nodes being subsequently merged while combining their respective connections with other nodes. This subsumption can also be understood as the walking node walking into the position of the chosen neighboring node. This basic merging of nodes is here understood from the perspective of a local \emph{subsumption} of the network topology. When performed successively along a random walk, the subsumption dynamics lead to a global condensation of the original network into a single node.

\begin{figure}
  \centering
     \includegraphics[width=0.5 \textwidth]{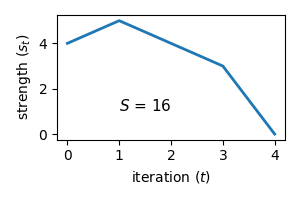}
   \caption{The strengths of the walking node in the example presented in Fig.~\ref{fig:walk} are shown in terms of the iteration $t$. The accumulated strength $S=16$, corresponding to the sum of the strength of the walking node for $t=0, 1, 2, 3, 4$, is also presented. The value of $S$ indicates the interaction between the walking node and its first neighbors accumulated along the complete subsumption dynamics.}\label{fig:signature}
\end{figure}

Given a complete subsumption of a network by a particular walking node, the strengths of the latter can be shown in terms of the iteration $t$, as illustrated in Figure~\ref{fig:signature}.

Observe that any subsumption of a complex network with size $N$ performed as described above will necessarily involve $N$ iterations $t=0, 2, \ldots, N-1$.

The subsumption of networks described above was motivated mainly by the intrinsically abstract interest and complexity of several of the concepts involved, including random walks, preferential dynamics, topological changes, and the interaction between topological and dynamical changes. In addition to this theoretical interest, the addressed systems also have the potential for modeling real-world situations corresponding to the gradual merging among companies, institutions, or administrative regions (e.g.~cities), among several other related situations. In addition, it is interesting to observe that other types of dynamics --- e.g.~involving agglomeration/coagulation of nodes --- can also be approached by the described concepts and methods.

The present work starts by describing the basic concepts and methodologies, and then presents and discusses several experimental results regarding the analysis of the subsumption of three types of networks, namely Erdős–Rényi (ER)~\cite{erdos1959random}, Barabási–Albert (BA)~\cite{barabasi1999BA}, and a geometrical network (GEO) corresponding to the Delaunay triangulation (e.g.~\cite{riedinger1988delaunay}) of a slightly perturbed version of an orthogonal lattice~\cite{benatti2023simple}, while considering random and two deterministic types of node walks.

\section{Basic Concepts and Methodology}\label{sec:methods}

Given a non-directed complex network having a size equal to $N$ nodes, it is possible to understand the number of connections of each node as a local measurement of the topology around each node. This measurement, which possibly corresponds to the most important local node property in a network, has been called \emph{node degree}. In the case of weighted complex networks, it is also possible to define the \emph{strength} of each node as corresponding to the sum of the weights of the respective existing links (e.g.~\cite{costa2007characterization,newman2018networks,barabasi2013network}).

Given a complex network, it is possible to perform random walks (e.g.~\cite{barber1970random,lovasz1993random,spitzer2013principles,xia2019random}) on it. There are several types of random walks, but one of the most frequently considered types consists of uniformly random walks performed by a single agent. In this case, after starting from a specific network node, the moving agent successively selects subsequent nodes with uniform probability among the existing connections.

In the present work, the distinct concept of \emph{node walk} is developed. This type of dynamics is characterized by moving a \emph{node} into other adjacent nodes, instead of moving an \emph{agent} from one node to another. The node is therefore said to perform a \emph{walk} throughout the network, incorporating the node and connections of each successively visited node, which implements a progressive subsumption of the network. A simple example of node walk is provided in Figure~\ref{fig:walk}.

Here, we considered both random and deterministic node walks. In the first case, the next node to be subsumed is chosen with uniform probability among the links of the walking node (first neighbors). Two types of preferential node walks are considered, to be the largest and smallest degrees among neighboring nodes.

Given the subsumption of a complex network by node(s) moving under specific rules, it becomes of particular interest to consider quantifications of the respectively obtained dynamics in terms of one or a small set of functionals (scalar measurements of the dynamic features). While there is virtually an infinity of such possible quantifiers, in the present work we consider the specific quantification, in terms of a single non-negative scalar value $S$, of the \emph{accumulated strength of the moving node}, obtained along the complete subsumption of the network. If the strength of the moving node at iteration $t$ is represented as $s_t$, we can write:

\begin{align}
  S = \sum_{t=0}^{N-1} s_t
\end{align}

The value $S$ is of particular interest because it summarizes in a single scalar measurement the total interaction of the moving node with other network nodes during the whole subsumption dynamics. More specifically, the value of $S$ is directly proportional to the average of the strength of the walking node along the whole walk. Thus, a high value of $S$ indicates that the moving node maintained a strong connection with other network nodes during its complete walk. Interestingly, the identification of motion rules leading to high (or low) values of $S$ constitutes a relatively intricate issue. For instance, in case the node always moves to its first neighbor having the largest degree, it would be expected that the strength of the moving node would increase more abruptly. However, if the topology of the specific network is such that most nodes with a high degree are interconnected, this specific rule could imply a quick stabilization (saturation) of the strength of the moving node.

Types of reasoning as considered above reveal that the resulting network subsumption, and particularly its $S$ measurement, can depend strongly not only on the walking rules but also on the topology of the covered networks. This provides an additional motivation for considering the total strength $S$ in the following analysis.

\section{Results and Discussion}\label{sec:results}

This section presents and discusses the experimental results obtained respectively to the subsuming of ER, BA, and GEO networks to the choice of the next node among the first neighbors of the walking node: (i) with uniform probability; (ii) having the largest degree; and (iii) having the smallest degree.

All experiments have been performed on complex networks with size $N=100$ and average degree $k=5.7$.

This section is divided into three subsections respectively to the analysis of the network subsumption by a single walking node analyzed locally at the node level and globally at the network level, as well as a preliminary study involving multiple walking nodes.

\subsection{Local Analysis}\label{sec:local}

The subsumption of complex networks by node walks is first approached at the local topological level corresponding to individual nodes and their respective links, with special attention given to the node degree and hierarchical degree with two neighborhoods (e.g.~\cite{da2006hierarchical}).

More specifically, specific networks of each of the three considered types (ER, BA, and GEO) are randomly chosen, which are then subsumed from each of the constituent nodes until the complete condensation of the original networks is achieved. In order to enhance statistical significance, 50 subsumptions are performed for each of the nodes to obtain respective signatures of the strength $s_t$ of the walking node in terms of the iteration $t$.

The obtained results have been characterized in terms of respective sets of individual accumulated strength signatures, as well as by scatterplots determined by the degree $k$ of the initial walking node \emph{versus} the obtained accumulated strength $S$. In this manner, it becomes possible to investigate the effect of the initial node degree on the respectively obtained overall dynamics, characterized in terms of the accumulated strength measurement. The Pearson correlation coefficient is also estimated for each obtained scatterplot, which provides a quantification of the linear relationship $\rho$ between the initial node degree and the accumulated strength. A particularly interesting question regards if it is possible to foresee the latter from the former property, which would be indicated by a relatively large value of $\rho$. Scatterplots have also been obtained for the hierarchical node degree considering two topological levels.

\begin{figure}
  \centering
     \includegraphics[width=1.0 \textwidth]{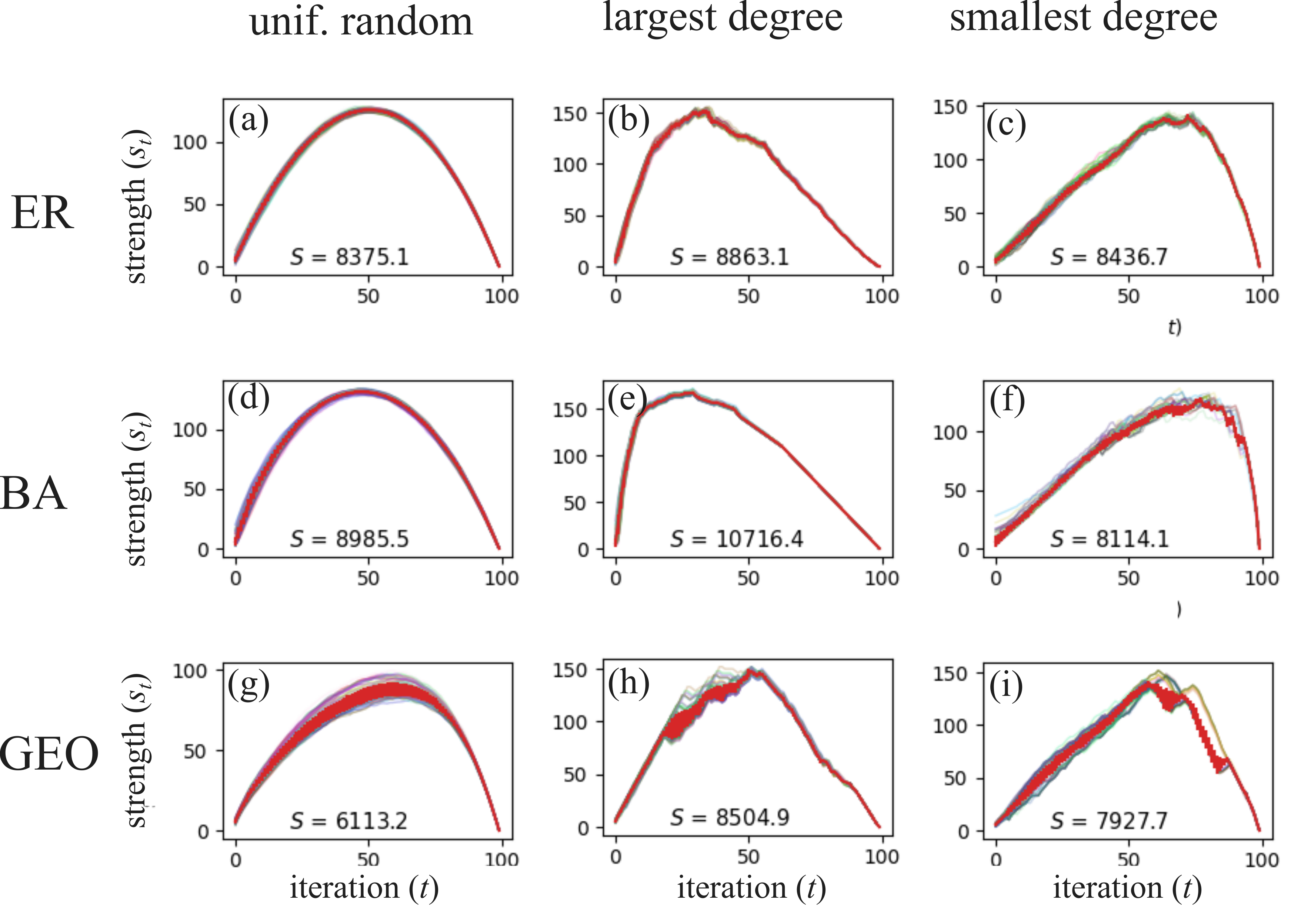}
   \caption{The set of signatures $s_t$ obtained for each of the nodes in ER, BA, and GEO networks under the three considered types of dynamics. The whole set of signatures in case are shown in random colors. The average $\pm$ standard deviation signatures in each case are shown in red. Mostly distinct shapes and dispersions have been obtained.}\label{fig:single_net_lavel}
\end{figure}

Figure~\ref{fig:single_net_lavel} presents the sets of individual accumulated signatures obtained for the ER, BA, and GEO structures under the three adopted subsumption dynamics.

Interestingly, though all obtained curves are characterized by a maximally peaked profile, markedly distinct signatures have been obtained for each considered combination of the type of network and subsumption dynamics, which are discussed as follows. Another surprising property of the obtained results is that, though the signatures depend strongly on the types of topology and subsuming dynamics, the very small standard deviations obtained in the case of ER and BA networks indicate that the signatures depend very weakly on the topology around each of the nodes of the network. 

\begin{figure}
  \centering
     \includegraphics[width=.85 \textwidth]{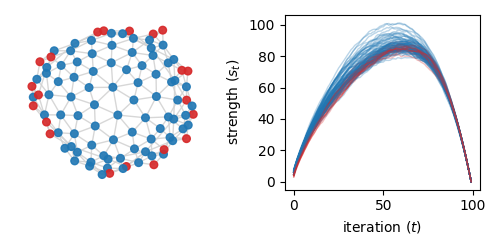} \\
     (a) \hspace{5cm} (b)
   \caption{The larger dispersion of the accumulated strength $S$ obtained in the case of the considered GEO networks is a consequence of the marked variation of node accessibility~\cite{travenccolo2009border,travenccolo2008accessibility,viana2010characterizing,benatti2022complex} and the presence of borders characterizing this type of network. The plate (a) shows a geographical network with $N=100$ nodes and its border nodes (in red) identified as the 20 nodes with the smallest node accessibility considering $h=3$ hierarchical levels. The signatures of accumulated walking node strengths obtained for uniform random choice of neighbors are shown in (b), with the colors being respective to that of the nodes in (a). The signatures obtained for the border nodes tend to present the smallest values of $S$.}\label{fig:border}
\end{figure}

The slightly large dispersion observed for the GEO model is mostly a consequence of this type of network having more pronounced borders, mainly as a consequence of not presenting the small world property. In other words, the border and more central nodes would tend to present distinct individual signatures for each of the network nodes. Figure~\ref{fig:border} illustrates this effect respectively to a GEO network with $N=100$ nodes.

The signatures obtained for the ER and BA networks under \emph{uniform random subsumption} resulted in the most bi-laterally symmetric cases, also presenting the smallest standard deviations. All other signatures are markedly asymmetric, though having varying shapes and standard deviations. In the case of the ER network, the respective symmetry and low dispersion are a consequence of its topological regularity, with most nodes having a similar topology. In the case of the BA structure, the symmetric and narrow distribution are also a consequence of the fact that the hubs are quickly reached along the subsumption, leading to similar subsequent dynamics. The signatures obtained for the GEO structure under uniformly random subsumption are also relatively narrow, but substantially more asymmetric than those obtained for the ER and BA cases (just discussed). In this case, it takes longer for the signature to reach its peak value, which is followed by a more abrupt decrease. This is a consequence of the presence of border nodes (having smaller degrees) in the GEO structure, which does not have the small world property, being characterized instead by a large average minimal topological distances between pairs of nodes.

The signatures obtained for the subsumption preferential to largest degree nodes all presented asymmetric profiles, especially in the ER and BA cases, being characterized by a fast increase of walking node strength $s_t$ followed by a more gradual decrease. The BA profiles resulted even more asymmetric than the ER and GEO counterparts for this type of subsumption dynamics. In the case of the BA network, this follows from the fact that the respective hubs are quickly reached, implying an abrupt increase of the respective walking node strength. The signatures obtained for the ER network occur for the same reason, with nodes of larger degree being obtained as a consequence of statistical fluctuations. As a consequence of having less marked hubs, the signatures obtained for the ER case increase less abruptly than those observed for the BA structure. The signatures obtained for the GEO network present a large dispersion around the average, as well as a tendency of the respective walking node strength to increase slower than when decreasing.

In the case of the subsumption taking place preferentially to the smallest degree, the results for the ER and BA structures basically correspond to the reversed signatures obtained for those same structures under preference to the largest degree, for similar reasons. Interestingly, this is not the case for the GEO network, which presented a highly asymmetric shape characterized by a slower almost linear increase followed by a sharp decrease. Another interesting difference between the signatures obtained in the case of the ER and BA networks for the two types of preferential subsumptions regards the fact that the case preferential to the largest degree yielded signatures characterized by smaller dispersion along the period from the beginning of the walk up to the peak of accumulated strength than the respective counterparts obtained for node walks preferential to the smallest degree.

\begin{figure}
  \centering
     \includegraphics[width=1.0 \textwidth]{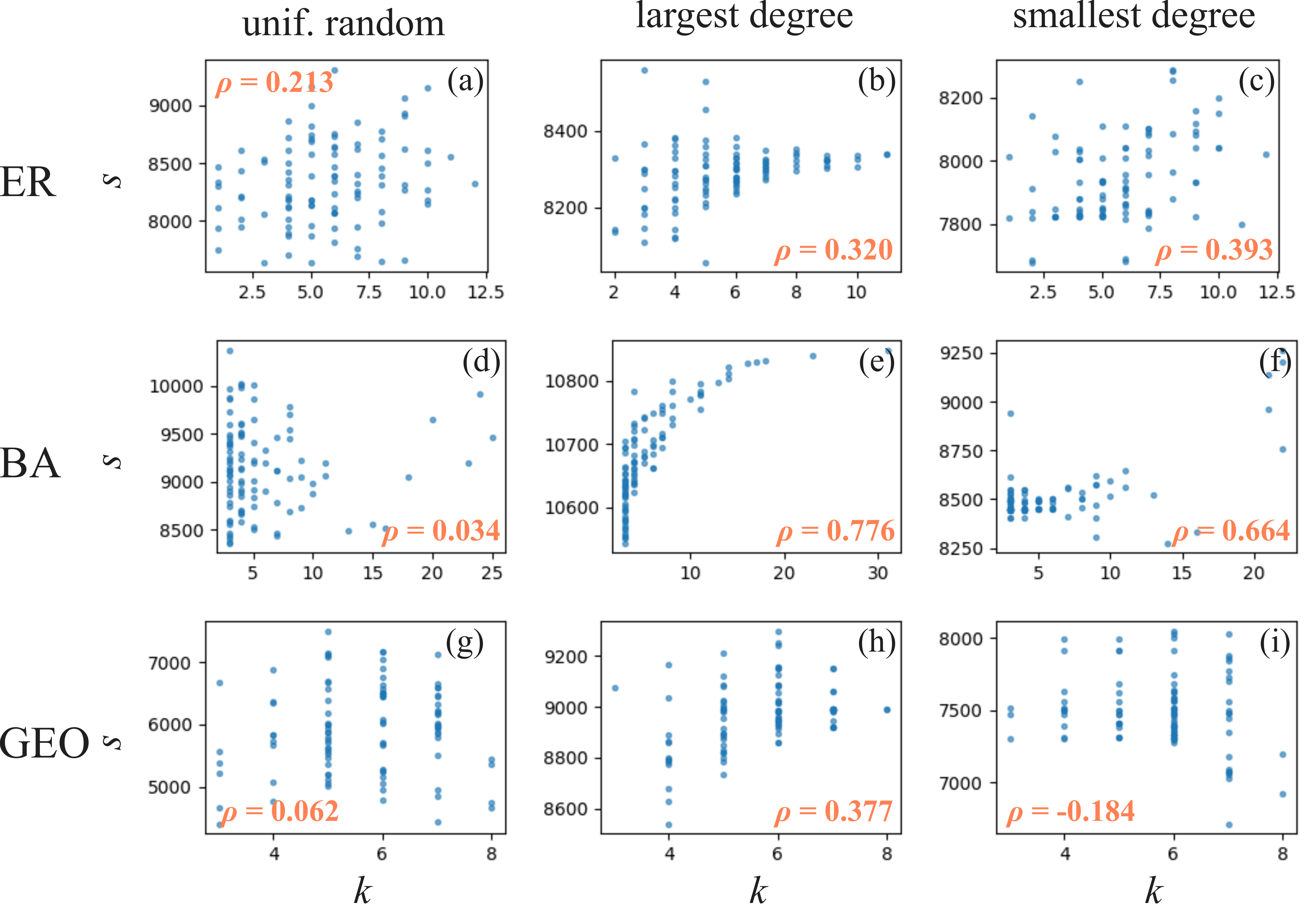}
   \caption{Scatterplots showing the relationship between the initial degree $k$ of the walking node and its total strength $S$. The Pearson correlation coefficients are respectively indicated. Three cases -- namely (c), (e), (f), and (h) -- resulted in larger correlation values. Observe, however, that the largest correlation does not correspond to a well-defined linear relationship.}\label{fig:node_level}
\end{figure}

Figure~\ref{fig:node_level} depicts the scatterplots obtained for respective samples of ER, BA, and GEO networks considering each of the three subsuming dynamics, namely uniformly random, as well as preferential to largest and smallest degrees of the adjacent nodes (first neighbors).

Several interesting indications can be inferred from the obtained results. First, we have that the obtained accumulated strengths are strongly dependent on both the type of network and on the adopted subsumption dynamics, as substantiated by the distinct distribution and ranges of $S$ in the scatterplots of Figure~\ref{fig:node_level}.

The interdependence between the two considered topological and dynamical measurements, namely $k$ and $S$, also results greatly dependent on the type of adopted network topology and subsumption dynamics, as suggested by the diversity of obtained distribution of points in Figure~\ref{fig:node_level}. 

Quite diverse Pearson correlation coefficients have also been obtained, varying from $-0.184$ to $0.776$. However, most of these coefficients have relatively small values, exceeding $0.35$ only for the cases (b), (c), (e), and (f). Interestingly, no statistically significant relationship has been observed for the case of uniform subsumption dynamics. In addition, relatively large correlations have been identified only for the ER and BA types of topologies, without including any of the GEO considered cases.

The relatively higher values of $\rho$ being found for the network types ER and GEO under the preferential walking dynamics are possibly related to the fact that these two types of topologies present a more pronounced degree of asymmetry. This is especially observed in the case of BA structures, which are characterized by scale-free node degree distributions, but also for the ER structures as a consequence of statistical fluctuations.

\begin{figure}
  \centering
     \includegraphics[width=1.0 \textwidth]{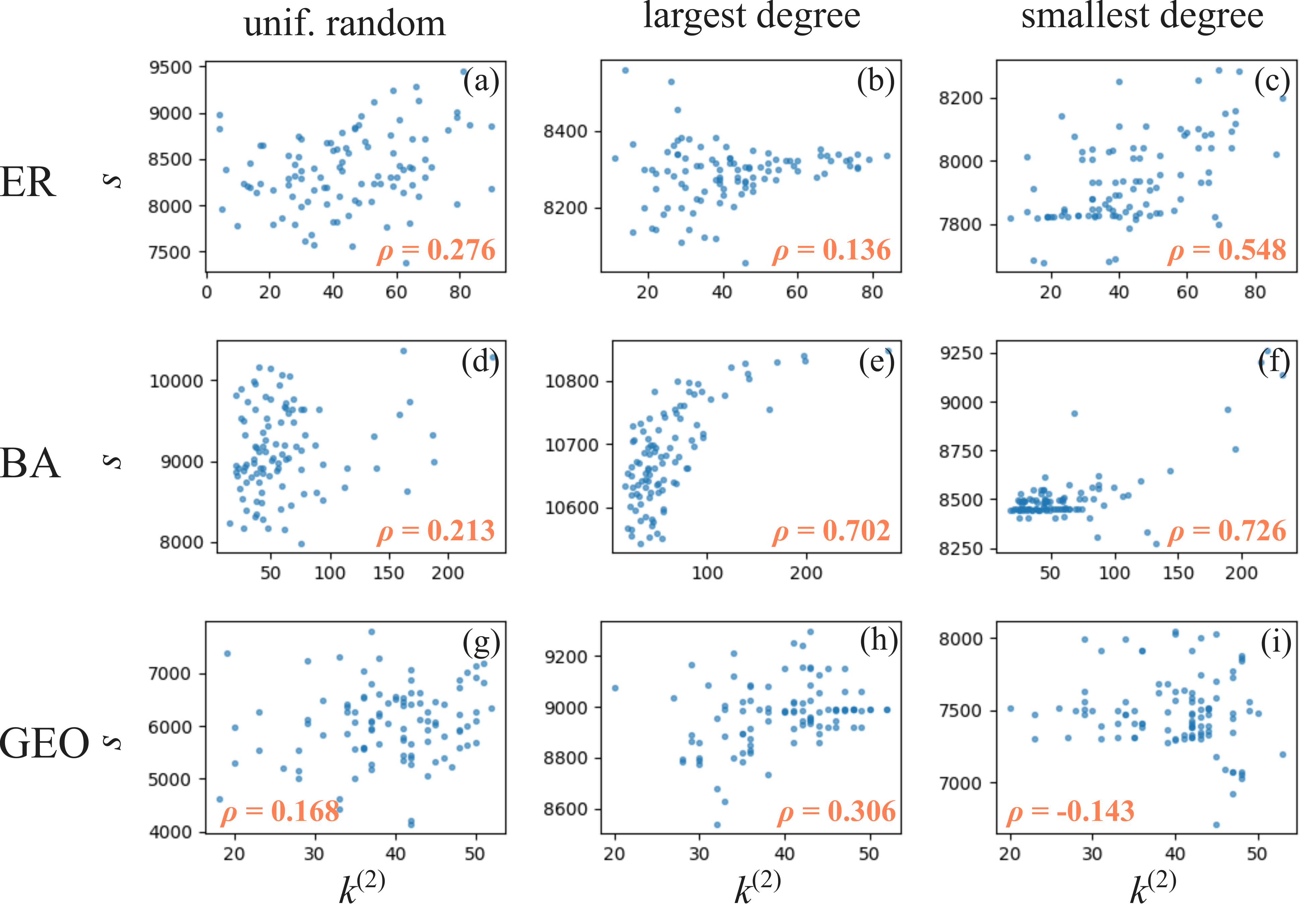}
   \caption{Scatterplots of the accumulated strength in terms of the cumulative node degree for the initial first and second neighbors of the walking node. Although moderately larger values of $\rho$ have been obtained in most cases, especially for (b), (e), (f), and (h), considerable dispersion of $S$ can still be observed in case (e) for small values of $k^{(2)}$.}\label{fig:node_level_2}
\end{figure}

Despite the relatively high values of $\rho$ obtained for the cases (e) and (f), there is still a substantial dispersion of $S$ for each fixed value of $k$. This effect is particularly intense in case (e), where a substantial dispersion of $S$ can be observed along the smallest values of $k$. This interesting effect suggests that the overall accumulated strength is determined not only by the initial degree of the walking node but also by additional network properties. In order to investigate this possibility further, scatterplots were also obtained respectively not to the initial node degree, but considering instead the initial cumulative first and second hierarchical degrees of the walking node, $k^{(2)}$. The results are shown in Figure~\ref{fig:node_level_2}.

The obtained results are characterized by moderately larger values of the Pearson correlation coefficient $\rho$, especially for the cases (b), (e), (f), and (h). This indicates that the cumulative node degree considering two hierarchical levels around the initial walking node do not allow a substantial increase in the relationship between this topological measurement and the dynamical property of strength accumulated along the whole subsumptions.

\subsection{Global Analysis}

\begin{figure}
  \centering
     \includegraphics[width=1.0 \textwidth]{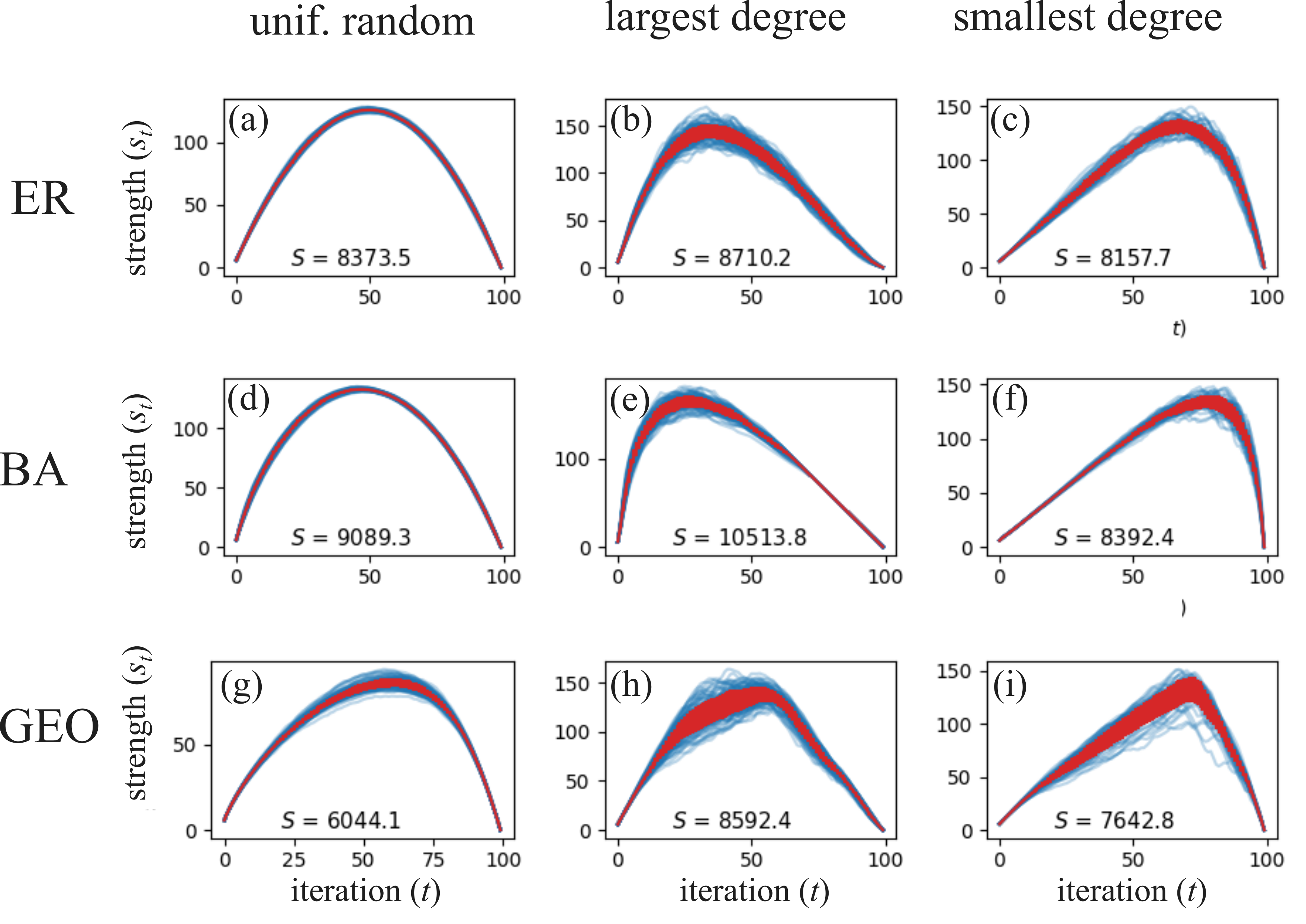}
   \caption{The set of signatures $s_t$ (in blue) obtained for each of the considered three types of topologies and subsumption dynamics, as well as the respective average $\pm$ standard deviation obtained curves (in red). Markedly distinct signatures have been obtained respectively to the several combinations of network types and subsumption dynamics.}\label{fig:net_level}
\end{figure}

In this section, the topic of subsumption of complex networks by node walks is approached at the global level of considering the average signature of the accumulated strengths for each of the adopted types of networks and subsumption dynamics. A total of 100 samples of networks were considered for each of the types ER, BA, and GEO. Random walks are started from each possible node, and the respective average and standard deviation of the accumulated strength in terms of $t$ ($s_t$) is then obtained. In the case of uniformly random dynamics, 50 random walks have been considered for each of the network nodes for the sake of enhanced statistical significance. Figure~\ref{fig:net_level} illustrates the obtained results.

As could be expected, the obtained signatures are closely similar to those obtained in the local analysis discussed in Section~\ref{sec:local}, with the main difference that relatively larger dispersions around the average signatures, as well as smoother shapes, have now been observed as a consequence of the consideration of not only a single network of each type but of respective ensembles of 50 samples.

The largest values of accumulated strength $S$ have been found for the BA network respectively subsumption preferential to the largest degree, followed by BA with uniformly random subsumption, and ER for dynamics preferential to the largest degree. This indicates that the wide range of node degrees observed in BA networks plays a decisive influence on the subsumption dynamics.

\section{Concluding Remarks}

Random walks of agents among the nodes of a complex network have been often considered as means for modeling and better understanding complex systems. In the present work, the possibility of having walks to be performed by nodes instead of agents. More specifically, one of the current neighbors of the walking node is selected according to some criterion, being incorporated with its links into the walking node, implementing a respective progressive \emph{subsumption} of the network. As a consequence, the network sizes decrease by one at each of the node movements. In addition, as the node walk progresses, its strength successively changes as a consequence of the incorporation of the visited node and respective links. This motivated the adoption of the instant waking node strength $s_t$, as well as it total accumulated strength $S$, as properties quantifying respective properties of the obtained dynamics.

The study of node walks has been motivated by two main possibilities: (a) to model related real-world systems including successive mergings of companies, towns, groups, etc.; as well as (b) to obtain an interesting complex system in which not only the topology and dynamics change in a coupled manner, but also which is potentially strongly dependent on specific network topologies and dynamics, allowing systematic studies of relationships between topological and dynamical changes in complex networks.

Three types of networks -- namely ER, BA, and GEO, as well as three types of dynamics -- uniformly random and deterministically preferential to the largest and smallest degree, have been considered. The experimental study of these combinations involved two main approaches, one local at the level of individual nodes, and another considering all all nodes in the networks. In the first case, specific attention was given to the relationship between the accumulated strength and topological properties of the initial walking node. In the latter case, the average $\pm$ standard deviation of signatures of the walking node strength were obtained and compared to one another.

Several interesting results have been obtained and discussed, including: (i) all obtained signatures are characterized by a peaked profile characterized by a growing stage followed by a decreasing phase; (ii) as expected, the subsumption dynamics depends strongly not only on the topology of the network, but also on the specific criterion adopted for the subsumption dynamics;(iii) at the same time, given a specific network, the signatures obtained for its distinct nodes tend to be markedly similar in the cases ER and BA; (iv) markedly diverse relationships have been observed between the accumulated strength and topological measurements of the topology around the initial walking node; (v) relatively large values of the Pearson correlation coefficient have been obtained for some of the considered cases, indicating some moderate possibility of predicting the accumulated strength from the topological properties around its initial configuration; and (vi) despite the relatively large correlation values obtained for some configurations, the considered initial topological properties of the walking node (degree and cumulative hierarchical degree) are not enough to fully determine the resulting subsumption dynamics. 

The reported concepts, methods, and results pave the way for several related future investigations. For instance, it would be interesting to consider other types of networks, including modular structures, as well as different types of subsumption criteria, including situations relaxing the requirement of the next node being adjacent to the current walking node (analogy to Levy flights, e.g.~\cite{shlesinger1986levy}). Another particularly interesting prospect consists of considering more than a single moving node, in which case the respective accumulated strengths can be compared and strategies for respective optimization be devised and evaluated. Yet another promising possibility consists of devising models for network growth motivated by the reversed sequence of events involved in the currently described subsumption approach. More specifically, new nodes could sprout from randomly selected existing nodes and establish connections mainly with the neighbors of that parent node, thus potentially establishing topological hierarchies in the so obtained networks.

\section*{Acknowledgments}
Alexandre Benatti thanks MCTI PPI-SOFTEX (TIC 13 DOU 01245.010\\222/2022-44). Luciano da F. Costa thanks CNPq (grant no.~307085/2018-0) and FAPESP (grant 2022/15304-4).

\bibliography{ref}
\bibliographystyle{unsrt}

\end{document}